\documentclass{jetpl}
\twocolumn

\lat

\usepackage{hyperref}

\newcommand{\gf}{\mathfrak{g}}
\newcommand{\af}{\mathfrak{a}}

\title{SLE martingales in coset conformal field theory}

\rtitle{SLE martingales in coset conformal field theory}

\sodtitle{SLE martingales in coset conformal field theory}

\author{Anton Nazarov$^{+*}$\/\thanks{e-mail: anton.nazarov@hep.phys.spbu.ru}}

\rauthor{Anton Nazarov}

\sodauthor{Nazarov}

\address{
  $^+$ Department of High-Energy and Elementary Particle Physics, 
  Faculty of Physics and\\ Chebyshev Laboratory,
  Faculty of Mathematics and Mechanics, \\ SPb State University
  198904, Saint-Petersburg, Russia}

\dates{12 May 2012}{*}

\abstract{
  Scharmm-Loewner evolution (SLE) and conformal field theory (CFT) are popular and widely used instruments to study critical behavior of two-dimensional models, but they use different objects. While SLE  has natural connection with lattice models and is suitable for strict proofs, it lacks computational and predictive power of conformal field theory.
  To provide a way for the concurrent use of SLE and CFT we consider CFT correlation functions which are martingales with respect to SLE.  We establish connection between parameters of Schramm-Loewner evolution on coset space and algebraic data of coset conformal field theory. Then we check the consistency of our approach with the  behaviour of parafermionic and minimal models. Coset models are connected with off-critical massive field theories and we discuss implications for SLE.} 

\PACS{05.50.+q, 11.25.Hf, 02.20.Sv}

\begin{document}
\maketitle

\section{Introduction}
\label{sec:introduction}

Schramm-Loewner evolution \cite{schramm2000scaling} is a stochastic process satisfying stochastic differential equation (\ref{eq:19}). A solution of  equation (\ref{eq:19}) can be seen as  probability measure on random curves (called SLE traces). SLE is useful for the study of critical behavior \cite{Cardy:2005kh,rohde2005basic}, since for many lattice models the convergence of  domain walls to SLE traces can be proved \cite{bauer20062d,schramm2006conformally}. Practical computations with SLE rely on Ito calculus while proofs use discrete complex analysis \cite{duminil2011conformal}.

Conformal field theory \cite{belavin1984ics} is formulated in terms of families of quantum fields which usually has no direct microscopic interpretation. On the other hand CFT has efficient computational tools such as Virasoro symmetry, Knizhnik-Zamolodchikov equations, fusion algebra etc and provides numerous predictions, such as celebrated Cardy formula \cite{cardy1992critical,smirnov2001critical} which gives crossing probability for critical percolation. 

Combination of these two approaches is immensely powerful, so it is important to find objects which can be studied from both points of view.

We present a new class of such observables in coset conformal field theory  -- correlation functions which are martingales with respect to Schramm-Loewner evolution with additional Brownian motion on factor-space of Lie group $G$ by subgroup $A$ \cite{2011arXiv1112.4354N}. We derive relations connecting SLE and CFT parameters. We then check the consistency of these results with the earlier results  for parafermionic observables \cite{santachiara2008sle}. 

All minimal unitary models can be obtained by coset construction. Coset structure leads to specific off-critical behavior. Massive excitations of coset CFTs are given by affine Toda field theories \cite{fateev1990conformal,eguchi1989deformations,hollowood1989rational}. Recent experimental study confirming this result attracted a lot of attention \cite{coldea2010quantum}. In conclusion we discuss possible connections of massive CFT perturbations with  off-critical SLEs (containing additional drift term  \cite{makarov2010off}).

\section{Martingale conditions}
\label{sec:mart-cond}

 Consider some critical lattice model on the upper half-plane $\mathbb{H}$ with the cut along a critical interface $\gamma_{t}$ (domain wall up to some length $t$). We denote this slit domain by $\mathbb{H}_{t}=\mathbb{H}\setminus \gamma_{t}$.  Assume that $\gamma_{t}$ satisfies restriction property and is conformally  invariant \cite{Cardy:2005kh}. Then  conformal map  $g_{t}:\mathbb{H}_{t}\to \mathbb{H}$ satisfies stochastic differential equation \cite{schramm2000scaling}:
\begin{equation}
\label{eq:19}
  \frac{\partial g_t(z)}{\partial t} = \frac{ 2}{g_t(z)-\sqrt{\kappa}\xi_{t}} ,
\end{equation}
where $\xi_{t}$ is the Brownian motion. The dynamic of the tip $z_{t}$ of critical curve $\gamma_{t}$ (tip of SLE trace) is given by the law $z_{t}=g_{t}^{-1}(\sqrt{\kappa}\xi_{t})$. 

For us it is convenient to use the map $w_{t} (z)=g_{t}(z)-\sqrt{\kappa}\xi_{t}$, so the equation \eqref{eq:19} becomes
\begin{equation}
  \label{eq:20}
       d w _{t}= \frac{2dt}{w_{t} }-\sqrt{\kappa}d\xi_{t}  
\end{equation}

Consider  $N$-point correlation function of boundary conformal field theory with boundary condition changing operators sitting at the tip of SLE trace and at the infinity:
\begin{equation}
  \mathcal{F}(\left\{z_{i},\bar z_{i}\right\}_{i=1}^{N})_{\mathbb{H}_{t}}=
\left<\varphi(z_{t}) \phi_{1}(z_{1},\bar z_{1}) \dots \phi_{n}(z_{n},\bar z_{n})
      \varphi(\infty)\right>,
    \label{eq:2}
\end{equation}
where $\phi_{i}$ are bulk primary fields with conformal weights $h_i$ and $\varphi(z_{t})$ is boundary condition changing operator.

Let $\gf$ be a (semisimple) Lie algebra of a Lie group $G$ and $\af$ -- Lie algebra of a subgroup $A\subset G$. Denote by $x_{e}$ embedding index for $\af\to\gf$. 

Primary fields in coset models are labeled by pairs $(\lambda,\eta)$ of weights for $\gf$- and $\af$-representations. For each $\lambda$ the set of possible weights $\eta$ includes those which appear in the decomposition  $L^{(\mu)}_{\gf}=\bigoplus_{\eta} H^{\lambda}_{\eta}\otimes L^{(\eta)}_{\af}$. (Some pairs are equivalent, see \cite{fuchs1996resolution,schellekens1990field}). We assume that boundary field $\varphi$ is also primary and is labeled by weights $(\mu,\nu)$.

  We use the conformal map  $w(z):\mathbb{H}\setminus\gamma_{t}\to \mathbb{H}$ to rewrite expression \eqref{eq:2} in the whole upper half plane:
\begin{multline}
  \mathcal{F}(\left\{z_{i},\bar z_{i}\right\})_{\mathbb{H}_{t}}=\\
  =\prod \left(\frac{\partial w(z_{i})}{\partial z_{i}}\right)^{h_{i}} 
  \prod \left(\frac{\partial \bar w(\bar z_{i})}{\partial \bar z_{i}}\right)^{\bar h_i}
  \mathcal{F}(\left\{w_{i}, \bar w_{i}\right\})_{\mathbb{H}}
  \label{eq:1}
\end{multline}
We can extend the definition of $\mathcal{F}$ to the whole plane by doubling the number of fields $\phi_{i}$ \cite{cardy2004boundary,cardy1984conformal} and by considering $w_{i},\bar w_{i}$ as independent variables. To simplify notations we will write $\mathcal{F}(\left\{w_{i}\right\}_{i=1}^{2N})$.

$G/A$-coset conformal field theory can be realized as a WZNW-model (with gauge group $G$) interacting with pure gauge fields of gauge group $A\subset G$ \cite{gawdzki1988g,figueroa89equivalence}. The action is written in terms of fields $\gamma:\mathbb{C}\to G$ and $\alpha,\bar\alpha:\mathbb{C}\to A$:
\begin{multline}
  \label{eq:24}
      S_{G/A}(\gamma, \alpha)=-\frac{k}{8\pi}\int_{S^2} d^2x\; {\cal K} (\gamma^{-1}\partial^{\mu}\gamma,\gamma^{-1}\partial_{\mu}\gamma)-\\
 - \frac{k }{24\pi} \int_{B}\epsilon_{ijk} {\cal K}\left(
    \tilde \gamma^{-1}\frac{\partial \tilde \gamma}{\partial y^i},
      \left[\tilde \gamma^{-1}\frac{\partial \tilde \gamma}{\partial y^j}
      \tilde \gamma^{-1}\frac{\partial \tilde \gamma}{\partial y^k}\right]\right) d^3y+\\
+
      \frac{k}{4\pi}\int_{S^2} d^{2}z \left(\mathcal{K}(\alpha, \gamma^{-1}\bar \partial \gamma)-\mathcal{K}(\bar \alpha, (\partial \gamma ) \gamma^{-1})\right.\\
      \left.+\mathcal{K}(\alpha,\gamma^{-1}\bar \alpha \gamma)-\mathcal{K}(\alpha,\bar \alpha)\right).
\end{multline}
Here we denote by $\mathcal{K}$ the Killing form of a Lie algebra $\gf$ corresponding to a Lie group $G$.

 If we fix $A$-gauge we have $G/A$ gauge symmetry. Then we add random gauge transformations to Schramm-Loewner evolution \cite{2011arXiv1112.4354N} similar to the case of WZNW-models \cite{bettelheim2005stochastic}.  Denote by $t^{a}_{i}$ ($\tilde{t}^{b}_{i}$) the  generators of $\gf$-representation ($\af$-representation) corresponding to the primary field $\phi_{i}$.

Now we need to consider the evolution of SLE trace $\gamma_{t}$ from  $t$ to $t+ dt$. First factor in the right-hand-side of equation (\ref{eq:1}) gives us
\begin{equation*}
  -\frac{2h_{i}}{w_{i}^{2}}\left(\frac{\partial w_{i}}{\partial z_{i}}\right)^{h_{i}}.
\end{equation*}
We denote by $\mathcal{G}_{i}$ the generator of infinitesimal transformation of primary field $\phi_{i}$:$d\phi_{i}(w_{i}) = \mathcal{G}_{i}\phi_{i}(w_{i})$. We normalize additional $\left(\dim\gf\right)$-dimensional Brownian motion as $\mathbb  {E}\left[d\theta^{a}\; d\theta^{b}\right]=\mathcal{K}(t^{a},t^{b})dt$. We also introduce the parameter $\tau$, which is the variance of this stochastic process. The generator of field transformation is equal to
\begin{equation}
  \mathcal{G}_{i}=\left(\frac{2dt}{w_{i}}-\sqrt{\kappa} d\xi_{t}\right) \partial_{w_{i}}+\frac{\sqrt{\tau}}{w_{i}}\left(\sum_{a:\mathcal{K}(t^{a},\tilde{t}^{b})=0}\left(d \theta ^{a} t^{a}_{i}\right)\right).
\label{eq:3}
\end{equation}
So we have fixed $A$-gauge by allowing random walk only in direction orthogonal to subalgebra $\af$. 

The differential of $\mathcal{F}$ should be zero due to martingale condition.  Ito formula is used to calculate the differential \cite{alekseev2010sle}. We need to include second order terms in  $\mathcal{G}_{i}$ since they contain squares of Brownian motion differentials $d\xi_{t},\; d\theta^{a}$ with the expectation values proportional to $dt$ ($\mathbb{E}[d\xi_{t}^{2}]=dt$):
\begin{multline}
d \mathcal{F}_{\mathbb{H}_{t}}=\\ \left(\prod_{i=1}^{2N}\left(\frac{\partial w_{i}}{\partial z_{i}}\right)^{h_{i}}\right)
\left(-\sum_{i=1}^{2N}\frac{2h_{i}dt}{w_{i}^{2}}+\left[\sum_{i=1}^{2N}\mathcal{G}_{i}+\frac{1}{2}
      \sum_{i,j}\mathcal{G}_{i}\mathcal{G}_{j}\right]\right)\mathcal{F}_{\mathbb{H}}\\=0
\label{eq:8}
\end{multline}
Substituting  definition \eqref{eq:3} we get 
\begin{equation}
  \left(-2 \mathcal{L}_{-2}+\frac{1}{2}\kappa \mathcal{L}_{-1}^{2}+\frac{\tau}{2}\left( \sum_{a} \mathcal{J}^{a}_{-1} \mathcal{J}^{a}_{-1}-
      \sum_{b}\tilde{\mathcal{J}}^{b}_{-1} \tilde{\mathcal{J}}^{b}_{-1}\right)\right)        \mathcal{F}_{\mathbb{H}}=0,
  \label{eq:4}
\end{equation}
with
\begin{eqnarray*}
  \mathcal{L}_{-n}=\sum_{i}\left(\frac{(n-1)h_{i}}{(w_{i}-z)^{n}}-\frac{\partial_{w_{i}}}{(w_{i}-z)^{n-1}}\right),\\ \mathcal{J}^{a}_{{-n}}=-\sum_{i}\frac{t^{a}_{i}}{(w_{i}-z)^{n}};\; \tilde{\mathcal{J}}^{b}_{{-n}}=-\sum_{i}\frac{\tilde{t}^{b}_{i}}{(w_{i}-z)^{n}}.
\end{eqnarray*}
This equation is equivalent to the following algebraic condition on the boundary state $\varphi(0)\left|0\right>$:
\begin{multline}
  \label{eq:7}
  \left<0\left|\varphi(\infty)\phi_{1}(w_{1})\dots\phi_{2N}(w_{2N})\right.\right.\\
  \left(-2L_{-2}+\frac{1}{2}\kappa L_{-1}^{2}+\frac{1}{2}\tau \left(\sum_{a=1}^{\dim\gf}J^{a}_{-1}J^{a}_{-1}-\sum_{b=1}^{\dim\af}\tilde{J}^{b}_{-1}\tilde{J}^{b}_{-1}\right)\right)\\
\left.\varphi(0)|0\right>=0.
\end{multline}
Since the set $\{\phi_{i}\}$ consists of arbitrary primary fields we conclude that 
\begin{multline}
|\psi>=\left(-2L_{-2}+\frac{1}{2}\kappa L_{-1}^{2}+\frac{1}{2}\tau \left(\sum_{a=1}^{\dim\gf}J^{a}_{-1}J^{a}_{-1}-\sum_{b=1}^{\dim\af}\tilde{J}^{b}_{-1}\tilde{J}^{b}_{-1}\right)\right)\\
\cdot\varphi(0)|0>
\end{multline}
is a null-state. Now we act on $\psi$ by raising operators to get equations on $\kappa,\tau$. Since in a coset theory commutation relations of full chiral algebra are
\begin{equation}
  \label{eq:18}
\begin{array}{ll}
  L_{n}= L_{n}^{\gf}-L_{n}^{\af}, &\\
  \left[L_{n}^{\gf},J^{a}_{m}\right]= -m J^{a}_{n+m},&\\
  \left[L_{n}^{\gf},\tilde{J}^{b}_{m}\right]=-m\tilde{J}^{b}_{n+m},&\\
  \left[L_{n}^{\gf},L_{m}\right]=[L_{n}^{\gf},L_{m}^{\gf}]-[L_{n}^{\gf},L_{m}^{\af}]=&\\
  \quad =(n-m)L_{m+n}+\frac{c}{12}(n^{3}-n)\delta_{m+n,0},&
\end{array}
\end{equation}
 it is more convenient to act with $L_{2}^{\gf}$ and $\left(L_{1}^{\gf}\right)^{2}$. 
 Applying $L_{2}^{\gf}$ we get
\begin{equation*}
  L_{2}^{\gf}\psi= \left(-8 L_{0}-c+ 3 \kappa L_{0}+\frac{1}{2}\tau (k \dim\gf-x_{e}k\dim\af)\right) \varphi_{(\mu,\nu)}=0
\end{equation*}
We use $L_{0} \varphi_{(\mu,\nu)}=h_{(\mu,\nu)} \varphi_{(\mu,\nu)}$, with the conformal weight  $h_{(\mu,\nu)}= \left(\frac{(\mu,\mu+2\rho)}{2(k+h^{\vee}_{\gf})}-\frac{(\nu,\nu+2\rho_{\af})}{2(k x_{e}+h^{\vee}_{\af})}\right)$ and the central charge $c=\frac{k\dim \gf}{k+h^{\vee}_{\gf}}-\frac{x_{e}k\dim \af}{x_{e} k+h^{\vee}_{\af}}$. This leads to the relation on $\kappa,\tau$:
\begin{equation}
  \label{eq:28} (3\kappa-8)h_{(\mu,\nu)}-c+\tau (k\dim\gf-x_{e}k\dim\af) =0.
\end{equation}
The second relation appears as a result of the $L_{1}^{\gf}$-action:
\begin{equation}
  \label{eq:21}
 -12 h_{(\mu,\nu)}+2\kappa h_{(\mu,\nu)} (2h_{(\mu,\nu)}+1) + \tau
(C_{\mu}-\tilde{C}_{\nu})=0,
\end{equation}
 where $C_{\mu}=(\mu,\mu+2\rho)$ and $\tilde{C}_{\nu}=(\nu,\nu+2\rho_{\af})$ are the eigenvalues of the quadratic Casimir operators $\sum_{a}t^{a}t^{a}$ and $\sum_{b}\tilde{t}^{b}\tilde{t}^{b}$ of Lie algebras $\gf$ and $\af$.
Relations \eqref{eq:28},\eqref{eq:21} are the necessary conditions for CFT correlation functions to be SLE martingales. 

From equations \eqref{eq:28},\eqref{eq:21} we immediately get $\kappa,\tau$ for each pair $(\mu,\nu)$ of $\gf$ and $\af$-weights. 

\subsection{Examples}
\label{sec:examples-1}
As an example consider  $\frac{\hat{su}(2)_{N}}{\hat{u}(1)_{N}}$-coset models which are equivalent to $Z_{N}$-parafermions.  The central charge is $c=\frac{3N}{N+2}-1=\frac{2N-2}{N+2}$. Conformal weights of primary fields with Dynkin indeces $(k,l)$ are $h_{(k,l)}=\frac{k(k+2)}{4(N+2)}-\frac{l^{2}}{4N}$.

Case $N=2$, $c=\frac{1}{2}$ corresponds to the Ising model, we have two non-trivial primary fields with conformal weights $h_{(2,0)}=1/2, h_{(1,1)}=1/16$. Substituting the field $\varphi_{(2,0)}$ into equations (\ref{eq:28},\ref{eq:21}) we get: $3\kappa-9+4\tau =0;\quad -3+\kappa+4\tau=0$. The solution is $\kappa=3, \tau=0$. For the field $\varphi_{(1,1)}$ the relations are $3\kappa-16+64\tau=0,\quad -64+9\kappa + 64\tau=0$, $\kappa=16/3, \tau=0$. So we have no additional motion for the Ising model and two possible values for SLE parameter $\kappa$  coinciding with the well-known results \cite{schramm2006conformally}.

For $N=3$ the parafermionic model central charge is $c=\frac{4}{5}$. Conformal weights are $h_{(0,0)}=0$, $h_{(0,2)}=h_{(0,-2)}=\frac{2}{3}\; \mathrm{mod}\; 1$, $h_{(2,0)}=\frac{2}{5}$, $h_{(2,2)}=h_{(2,-2)}=\frac{1}{15}$.  The corresponding values of $\kappa,\tau$ are: $(\frac{208}{25},\frac{242}{225}), (\frac{10}{3},0),(\frac{80}{19},\frac{14}{171})$. As it  was mentioned in \cite{santachiara2008sle} the field $\varphi_{(2,0)}$ with the conformal weight $h_{(2,0)}=\frac{2}{5}$ constitutes the $Z_{3}$-singlet, so  an additional random walk does not appear and $\tau=0$. The form of equations (\ref{eq:7}) is similar to that of \cite{santachiara2008sle}, but the normalization of $\tau$ differs.

It is easy to see that for the $\frac{\hat{su}(2)_{N}\oplus \hat{su}(2)_{1}}{\hat{su}(2)_{N+1}}$-coset realization of minimal unitary models with $c=\frac{3N}{N+2}+1-\frac{3(N+1)}{N+3}=1-\frac{6}{(N+2)(N+3)}$ the system of equations (\ref{eq:28},\ref{eq:21}) is always consistent for $\tau=0$ and we get a standard connection between the SLE-parameter $\kappa$ and the central charge $c=\frac{(6-\kappa)(3\kappa-8)}{2\kappa}$.

\section{Conclusion and outlook}
\label{sec:outlook}

In  \cite{bettelheim2005stochastic} a connection between WZNW-models and Schramm-Loewner evolution with additional Brownian motion on group manifold was established. The authors also stated a problem of a possible connection of SLE parameters for martingales in WZNW, coset and minimal models. In present letter we used the method proposed in \cite{alekseev2010sle} to obtain necessary conditions on SLE martingales. This method allowed us to compare our results with parafermionic results presented in \cite{santachiara2008sle,picco2008numerical}.

The coset structure of minimal models is manifest in field theory perturbed by an external magnetic field \cite{fateev1990conformal,eguchi1989deformations,hollowood1989rational}. This theory is supported by experimental data \cite{coldea2010quantum}. Relations between correlation functions in coset theory and SLE observables can be a starting point in studies of domain walls in lattice models away from the  critical point.

Massive perturbations of $G/A$-coset theory are realized as an affine Toda field theory and are classified by simple roots of the Lie algebra $\gf$. Affine Toda field theory can be obtained by adding perturbation term to  the action (\ref{eq:24}):
\begin{equation}
  S_{\text{pert}}=S_{G/A}(\gamma,\alpha)-\frac{k\lambda}{2\pi}\int {\cal K} (\gamma T, \gamma^{-1} \bar T),
\end{equation}
where $T,\bar T\in \gf$ are specially chosen Lie algebra elements \cite{bakas1996lagrangian,hollowood1995massive,park1994deformed}. 
The perturbation leads to an insertion of certain primary field in all the correlation functions  \cite{hollowood1989rational}. It was shown that massive off-critical SLEs have additional drift term in driving Brownian motion \cite{makarov2010off,bauer2009off}.
The question is whether the interaction of the perturbing primary field with the $\tau$-term of the equation (\ref{eq:4}) leads to the same contribution as the addition of a massive drift to SLE.

In the forthcoming studies we will address this question and compare  massive perturbations of coset models with the numerical studies of domain walls in the Ising model perturbed by a random Gaussian external field \cite{stevenson2011domain}.

\section*{Acknowledgements}
\label{sec:acknowledgements}
 I am grateful to A. Bytsko, K. Izyurov, D. Chelkak and R. Santachiara for the discussions on SLE and WZNW models and to V.D. Lyakhovsky for the careful reading of the text.  This work  is supported by
the Chebyshev Laboratory (Department of Mathematics and Mechanics,
Saint-Petersburg State University) under the grant 11.G34.31.0026
of the Government of the Russian Federation.

\bibliography{bibliography}{}
\bibliographystyle{iopart-num}

\end{document}